# Intelligence gathering by capturing the social processes within prisons

Vassilis Kostakos and Panos A. Kostakos

Department of Computer Science, Department of European Studies and Modern Languages
University of Bath, Bath BA2 7AY, UK
`{v.kostakos, p.kostakos}@bath.ac.uk`

**Abstract**. We present a prototype system that can be used to capture longitudinal socialising processes by recording people's encounters in space. We argue that such a system can usefully be deployed in prisons and other detention facilities in order help intelligence analysts assess the behaviour or terrorist and organised crime groups, and their potential relationships. Here we present the results of a longitudinal study, carried out with civilians, which demonstrates the capabilities of our system.

## 1. Introduction

In this paper we argue that the deployment of pervasive technology in detention facilities can provide intelligence in relation to the activities of terrorist and organised crime groups, as well as their emerging relationships. Evidence suggests that detention facilities are increasingly becoming fora where terrorists and organised criminals establish channels of communication and co-operation, and more importantly recruit new members. Here we argue that the systematic capturing and analysis of the social processes within detention facilities can enhance intelligence and law enforcement agencies' understanding of the groups' operation and behaviour.

To demonstrate the type of data that can be obtained from detention facilities, we present the results of a longitudinal study we carried out in the City of Bath, UK, involving civilians who socialised in various locations across the city. Here we discuss how we were able to automatically capture and analyse data on people's encounters, and we present the results of our analysis. While our study did not take place in an actual detention facility, nevertheless it did take place in a real world setting and, as such, provides useful insights into how pervasive technologies may be utilised within detention facilities.

## 2. Terrorism and organised crime

Since the end of the Cold War era, the international community's public and scholarly interest has shifted towards security issues related to the rise of transnational criminal and terrorist networks that are perceived to threaten national and international security and stability [UN; Shelly 1995]. The nature of the threats poised by organised crime and terrorism renders their containment by state actors extremely difficult and problematic. Prime examples of this difficulty are newly established states in various troublesome regions such as the Former Yugoslavia and the Soviet Union, where states have been unsuccessful in managing effectively the activities of criminal organisations. This has permitted local Mafia groups, which claim huge profits from illicit markets, to fill political vacuums and develop symbiotic relationships with state institutions [Williams, 2000].

Terrorist organisations and groups have been dealt with much more proactively. The measures intended to combat terror activities often require exceeding retaliation by means of military intervention and in general methods that have been argued undermine human right and civil liberties. Examples of these measures include pre-emptive strikes, the establishment of Guantanamo Bay detention centre, alleged Rendition flights and increasingly draconian legislation in both US and UK.

These emerging security challenges are taking place in a globalised environment were distant social systems are becoming increasingly interconnected and interdependent. Migration flows from the East to the West, and from the South to the North, facilitated by improvements in communication and transportation technologies, are contributing to the growth of heterogeneous and multiethnic societies. This increased pace of physical and electronic interconnection between actors from distant social systems has contributed to a rise of weak social ties. According to Granovetter's hypothesis [1973], the establishment of weak ties or bridges amongst previously isolated groups,enhance the effectiveness of these actors' organisational structures and thus facilitates the materialisation of their goals. Both organised crime and terrorism has been argued to adhere to the same social principles [Chambliss 1971; Cohen 1977; Lombardo 1994; Williams 1998,2001; Kleemans & Van de Bund 1999].

As a result, over time terrorist organisations have developed resilience, and have been able to establish intricate channels of communication in order to improve and learn from their previous mistakes. Most significantly, terrorist networks are increasingly becoming able to study the operational behaviour of security forces, and frequently engage in counter intelligence practices. At the same time, the increased embeddedness of these terrorist networks within society makes it easier and more likely to recruit and radicalise through propaganda civilians of various social classes and professions. Consequently, valuable intelligence that could be fed to ongoing investigations is very likely to emerge from unconventional locations and sources, which the security apparatuses underestimate or cannot monitor effectively.

## 2.1. Prisons as a source of intelligence

The changing structure of the prison population in many European countries and the high number of foreign inmates [Council of Europe] increases the chances of Islamic militants being imprisoned along with "well-connected" criminals and individuals vulnerable to indoctrination methods and susceptible to radicalisation. Moreover, the criminal networks that are established in prisons offer significant financial and logistical resources, which can facilitate large-scale terrorist attacks [Shelley et al., 2005]. These conditions have raised concerns, already expressed by state officials[1], with regards to the increased possibilities facing detainees in various detention facilities. Sadly, these concerns have been verified through a number of case studies.

For instance, in 2001 Jose Emilio Suarez Trashorras was jailed in a Spanish prison for drug related offences. Whilst imprisoned, Trashorras established regular contact with Jamal Ahmidan who was serving time for a petty crime. Both individuals embraced radical Islamic fundamentalist ideas within the prison and were recruited in the Takfir wa al-Hijra group, a Moroccan terrorist groups linked with al-Qaida [Cuthberson, 2004]. Following their release, Ahmidan became the leader of the terrorist cell that conducted the Madrid bombing. In a drugs-for-bombs exchange with a third party, Trashorras provided the cell with explosives for the 13 backpack bombes that killed 191 people and injured hundreds.

Another vivid example of the role of detention facilities as recruitment pools of terrorist groups has been the case of the Martyrs of Morocco terrorist cell. This group was composed of 18 north African immigrants who were radicalised and recruited whilst serving a prison sentence for minor offensives including weapon possession, document fraud and robbery [McLean, 2004]. According to official sources, the leader of the Martyrs of Morocco cell co-ordinated an attack to bomb the national high court in Madrid and for that pursuit attempted to purchase 500 kilograms of explosive materials,[2] but was detained before carrying out the attack. It is worth noting that the Martyrs of Morocco cell was connected through prisons with the spiritual leader (emir)

---

[1] Testimony of John S. Pistole, Assistant Director, Counterterrorism Division, FBI Before the Senate Judiciary Committee, Subcommittee on Terrorism, Technology, and Homeland Security October 14, 2003: "Terrorist Recruitment in Prisons and The Recent Arrests Related to Guantanamo Bay Detainees". See http://www.fbi.gov/congress/congress03/pistole101403.htm

[2] Madrid ABC: Spanish Judge Orders Remanding of Islamists Involved in Bomb Plot (24-Oct-2004)

of the Madrid bombing and with members of the ETA terrorist group [Bar et al., 2005; Haahr-Escolano, 2004].

## 3. Pervasive technology for detention facilities

As part of our research, we prototyped a pervasive system that captures longitudinal socialising processes by recording and analysing people's encounters in space. To achieve this, we utilised Bluetooth technology, typically found in mobile devices. Bluetooth technology has a characteristic that renders it appropriate for studying people's encounters. In contrast to the wireless signals emitted by typically static WiFi access points, the signals emitted by Bluetooth devices map very closely to the movements of people around the city, which in turn are a unique indicator of encounter and socialising. In previous work, we found that approximately 7.5% of observed pedestrians had discoverable Bluetooth devices [O'Neill et al., 2006]. This number most certainly varies between different cities, but still it shows that a considerable portion of the public was recorded using our method.

Our basic setup, replicated across 4 sites, involved installing a computer that constantly recorded the presence of nearby Bluetooth devices within a 10-meter range (Figure 1). This data enables us to correlate pedestrian movements with Bluetooth device movements, providing baseline data about the penetration of Bluetooth into city life. On the right side of Figure 1 we see that for each unique device (i.e. person), we are able to capture *sessions*, defined as the points in time when each person was in close range of the scanner (indicated as yellow horizontal bars). Subsequently, we are able to detect *encounters* (indicated as links between the sessions), which we define as overlapping sessions. In other words, an encounter takes place when two people are in the same place at the same time.

In our study we considered four locations, which we shall refer to as

- campus

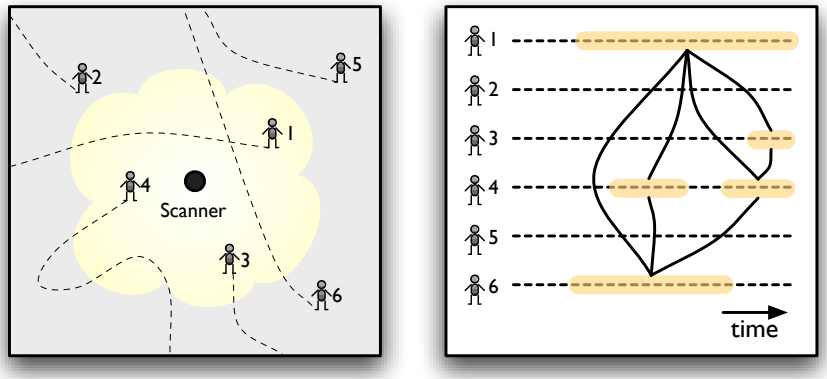

**Figure 1**. Left: At each scanning location, our computer uses Bluetooth to monitor the presence of mobile devices within an approximate 10 meter radius. Right: Each recorded device is allocated its own timeline (dotted horizontal lines). Using data from our scanners, we can plot each device's visit sessions (yellow bars). Overlapping sessions are identified and linked (solid lines), thus indicating encounters.

- street

- pub

- office

The first two locations are outdoor pedestrian streets, one on our campus and one in the city of Bath, both of which connect open spaces and can be thought of as pedestrian gateways. The latter two are indoor locations where visitors typically spend some time in them. The pub is open to anyone over the age of 18, while the office is a secure location where only employees and their visitors have access.

We should point out that the nature of Bluetooth technology mitigates against extreme accuracy of location. The 10-meter range of our Bluetooth scanner reached beyond walls, and in adjacent offices. Effectively, if our scanner picked up a Bluetooth device, there is no way of knowing if that device was on the street, or in any of the offices. Despite this, on aggregate level we still get quite distinctive patterns of data between the first two and last two locations, as we describe in the next sections. This is because the great majority of devices our scanners picked up was indeed on the street (for the first two locations). During a six month study of our prototype, we captured approximately 10,000 unique devices. In the following sections we describe in detail the data we captured and the analyses we carried out.

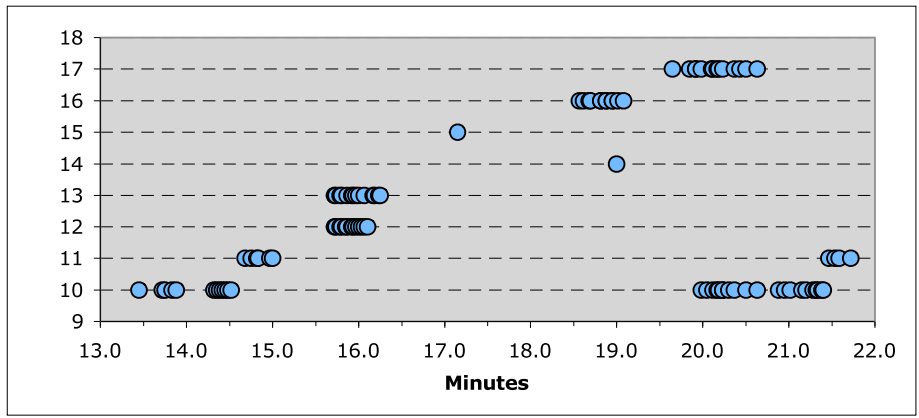

**Figure 2**: A timeline visualisation of our Bluetooth gatecounts. Each device is given its own timeline (dashed lines) and each discovery event is plotted as a circle on the timeline.

## 4. Data & Analysis

The method we used to scan for Bluetooth devices generates discrete data about the presence of devices in the environment. A visualisation of our raw data, which we have termed *timeline*, can be seen in Figure 2. Here, each dot represents a discovery event, i.e. a point in time (x-axis) when our Bluetooth scanner picked up a specific device in the environment. By applying filters, we can see that, for example, device 16 was present in the environment between approximately 18.5 minutes and 19.5 minutes.

To study the patterns of co-presence in our data, we first need to identify instances where two or more devices were present at the same place and the same time. For example, in Figure 2 we see that devices 12 and 13 encountered each other. We developed filters that analysed our data and gave us instances of devices encountering each other at each of the four locations in our study. These initial results took the form of records: <device1_id, device2_id, location>

At this stage in our analysis we had a long list of such records, describing which devices encountered each other and in which location. For example, in Figure 1 we see that devices 12 and 13 encountered each other at 15.5 minutes and were together for approximately 1 minute. This list of encounters is a textual representation of the patterns of encounter across our four locations. To further study the patterns and structure hidden within this list, we transformed it to four social network graphs, one for each location. Assuming that each device from our dataset becomes a node in the social graph, then the list of encounters indicates which nodes are connected. Proceeding in this manner, we generated four social network graphs, one for each location.

For illustration purposes, in Figure 3 we show the graph from the pub location in our study. In this graph, each device is represented as a node in the graph, and connected nodes indicate that these devices encountered each other at some point. We see that most devices are linked to the main core, whilst some devices are islands. The latter indicates cases where a device was seen only by itself and never in the presence of others. Additionally, the size of nodes represents the total amount of time that a device has spend in this location, while the colour of the nodes (blue to red) indicates the betweenness of a node (from 0 to 1 respectively).

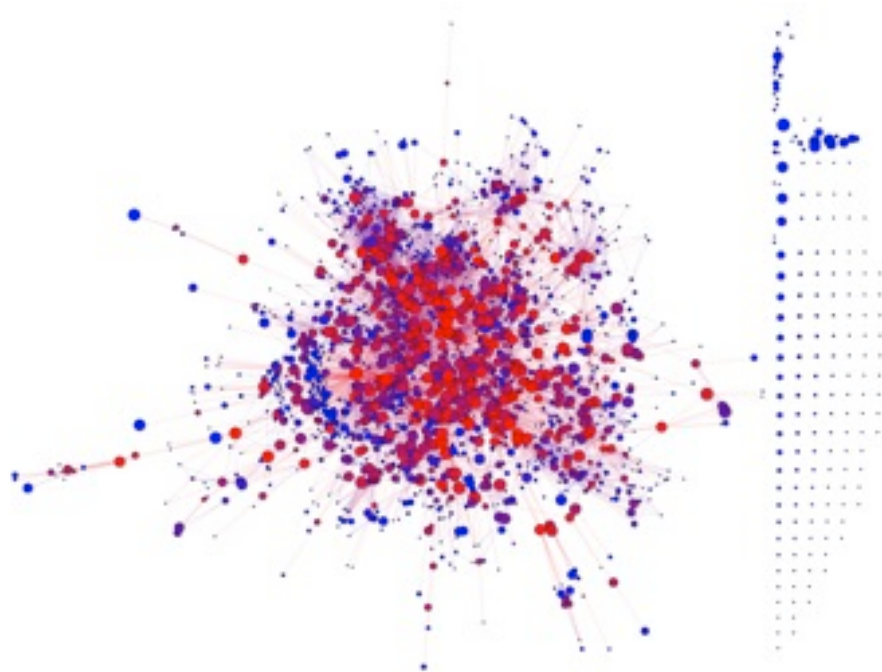

**Figure 3**: A graph visualisation of the encounters that we recorded at one of the locations in our study.

One of our initial observations was that due to the sheer number of nodes in the graphs, the visualisations themselves helped little in analysing our data, due to visual clutter. However, by transforming our data into graph form, we were able to run a number of well-established analysis algorithms using existing software (e.g. Pajek, Ucinet). Specifically, we analysed each of our four graphs in terms of

- Degree centrality, calculated as the number of neighbours of each node.

- Closeness centrality (access), calculated for any given node as the number of nodes (minus 1) divided by the sum of all distances between the node and every other node.

- Betweenness centrality (control), calculated for any given node as the proportion of shortest paths between all pairs of nodes that include this node.

- Distance, calculated as the probability that the shortest path between a random pair of nodes will be $1, 2, 3$, etc.

The degree and closeness centrality are measures of the reachability of a node within a network, and describe how easily information can reach a node. Betweenness centrality indicates the importance of a node, and the extent to which it is needed as a link in the chains of contacts that facilitate the spread of information within the network. Effectively, the centrality measures we focused on can indicate each individual's role, or potential, within the observed social structure.

|                                   | Campus | Street  | Pub    | Office |
|-----------------------------------|--------|---------|--------|--------|
| Unique devices                    | 1162   | 8450    | 4175   | 329    |
| Largest core                      | 1028   | 2738    | 4036   | 318    |
| 2nd largest core size             | 2      | 4       | 2      | 1      |
| Edges in largest core             | 6434   | 5060    | 23919  | 2419   |
| Density                           | 0.5%   | 0.007%  | 1.4%   | 2.2%   |
|                                   |        |         |        |        |
| Network Degree Centralisation     | 0.43   | 0.51    | 0.68   | 0.73   |
| Network Closeness Centralisation  | 0.49   | 0.55    | 0.66   | 0.65   |
| Network Betweenness Centralisation| 0.36   | 0.65    | 0.57   | 0.27   |
|                                   |        |         |        |        |
| Max degree                        | 454    | 1394    | 2758   | 246    |
| Average degree                    | 12.26  | 3.70    | 11.85  | 15.21  |
| Max distance (diameter)           | 6      | 10      | 9      | 4      |
| Average distance                  | 2.72   | 2.96    | 2.44   | 2.04   |
| Average clustering coefficient    | 0.50   | 0.32    | 0.68   | 0.82   |

**Table 1**: Metrics for each of our four graphs.

## 5. Results: capturing social processes

To gain an overview of the structural properties of the graphs representing encounter, we calculated the metrics shown in Table 1. For each of our locations we calculated the number of unique devices that were recorded by our Bluetooth scanner, the size of the largest core in the encounter graphs, the number of edges in the largest core, the density of the largest core as well as the size of the 2nd largest core. We also calculated some generic centrality measures for each of the largest cores: network degree, closeness and betweenness centralisation. Finally, we measured the maximum and average degree of each graphs, the longest shortest-path distance in each of the graphs, as well as the average shortest-path distance.

In addition to the above metrics, for each of degree, closeness and betweenness centrality measures we generated ranked log-log plots. To do this we attached a value (either degree, closeness or betweenness) to each node in the graphs (only the core), and then sorted this list in descending order. We then plotted the sorted lists, resulting in three sets of graphs (degree, closeness, betweenness) for each of our four gates. Additionally, we generated a fourth set of graphs, based on the probable distance between any randomly selected pair of nodes. These graphs are shown in Figures 4 to 7.

### 5.1. Structural measures

Our results indicate that the data captured by our prototype is far from random. On the contrary, across the four locations of our study we identified homogeneous patterns and comparable underlying temporal behaviour. To demonstrate this, here we focus our discussion on the various properties of the social graphs that we listed in Table 1. The way we captured and analysed our data prohibits us from directly inferring intelligence for each of the social networks. However, by comparing the properties the social graphs across our four locations we can begin to draw a picture of the communities that inhabit those locations. Also, it is important to keep in mind that in our observations of the four locations the only parameter we changed was the location itself: the hardware, software and algorithms we used to derive our results are identical for all locations. Although it can be argued that our data are affected by a number of further variables, we consider those as part of the location and the environment.

A notable feature of the graphs is their size. As we expected, the city street had the most "visitors", followed by the pub, the campus and the office. This is quite representative of the populations inhibiting each of the locations, since the street is open to everyone, thus likely to get lots of distinct visitors. The pub is also open to everyone (over 18) and again has a large population of potential visitors. The campus, on the other hand, is mostly visited by students and staff, which amount to about 15,000 students and staff (while the population of Bath is about 86,000). Finally, the office is a secure area where only employees have access, thus a small population of potential visitors.

It is interesting to note, however, that the social network of the street consists of about 2/3 islands, with the core consisting of about 1/3 of the devices.

Looking at Table 1 we see that the campus has a much higher density than the street. This indicates that there are more static devices on the campus, such as computers or employees phones, which are likely to act as hubs which connect to the core those single devices that go past in the environment. This is something we can verify from Figure 4, where we see the street graph has a few well connected hubs but then falls quite sharply, as opposed to the campus where there are many more nodes with degree between 100 and 5.

It is interesting to note that both locations where the public can go, the street and the pub, have quite large max-degree (1394 and 2758), yet average degree is much smaller on the street than the pub (3.70 and 11.85). In fact, in Figure 4 we see that the pub completely outperforms the street in terms of degree. This is due to the fact that most people in the pub are co-present, thus they get linked together. In other words, a visit in the pub can give someone much more opportunity for copresence than a visit in the street. This is something we expect, as it is the primary purpose of a pub. Also, we should note that in the pub there are certain devices with extremely high degrees, which we believe are attributed to members of staff or regular customers. These act as central hubs that bring together all the customers of the pub into the central core of the social graph. The same is true in the office, where a number of devices have a relatively high degree, indicating that these people come in frequent contact with others.

### 5.2. Network centrality measures

In general, across the four locations the "tightness" of the communities varies. Specifically, the office and the pub have shorter average distances between their members (2.04 and 2.44 in Table 1 respectively), and we also see in Figure 7 that the probability curves of these two locations are shifted to the left. This is further enhanced by the relatively high density of the pub and the office, which indicates more interactions between the members of the community.

Another interesting point to note is that although the pub has quite a tight and dense population, it has large diameter (9), which is also true of the street (10). Yet, the pub has a smaller average distance (2.44) as opposed to the street (2.96). Coupled with the density measures, we can describe the pub's network as a large central core, while the street's network more closely resembles a small core with a number of branches and additionally a large number of islands.

Considering the network centralisation measures we can make more inferences about the overall structure of the social networks. These measures range from 0 to 1 and indicate a similarity to a perfect linear-shaped network (0) or to a perfect star-shaped network (1). This is calculated for each of degree (DC), closeness (CC) and betweenness (BC). The office scores high on DC and CC indicating that some nodes can be reached more easily than others, yet BC is low, indicating that all nodes are more or less equally important in terms control and communication. The opposite is true of the pub, where high DC and CC are coupled with high BC. This indicates that there are certain nodes in the pub that act as hubs of communication and control (most likely the members of staff or regular customers). Comparing the campus and street in terms of centralisation measures also yields interesting insights. Both have similar levels of DC and CC, but the campus has low BC while the street has high BC. This indicates that on the street there are a few important nodes, while on campus the nodes are more equal.

### 5.3. Cumulative distribution measures

We now consider the graphs shown in Figures 4 to 7, which we found much more useful than a visualisation of the social networks themselves. A really interesting observation is that although in each of the 4 graphs the lines have similar shape, the subtle differences are crucial pointers as to the effect of space on encounter. For instance, the variation in how sharply the values fall is a useful indicators, along with the overall steepness of the graphs.

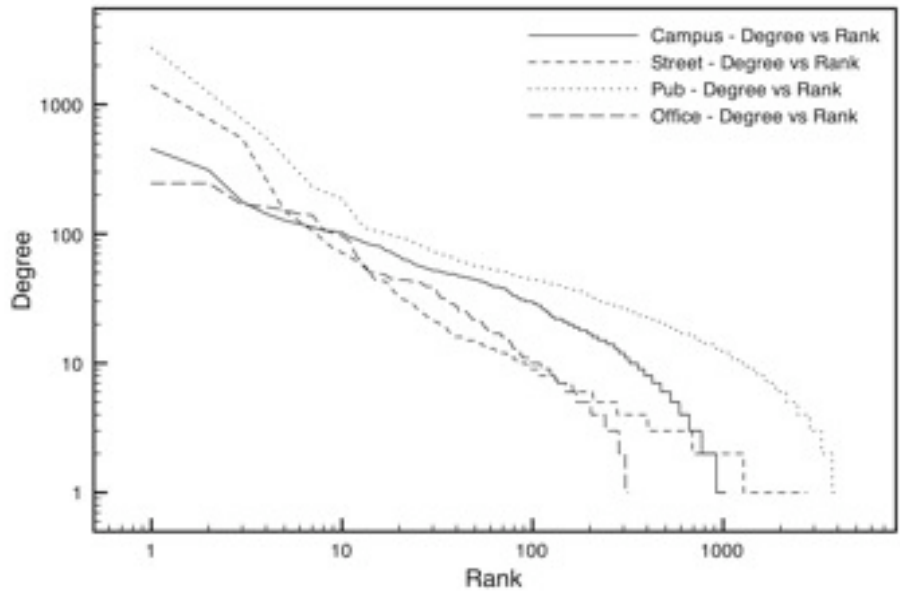

**Figure 4**: Ranked log-log plots of degree for each of our four locations.

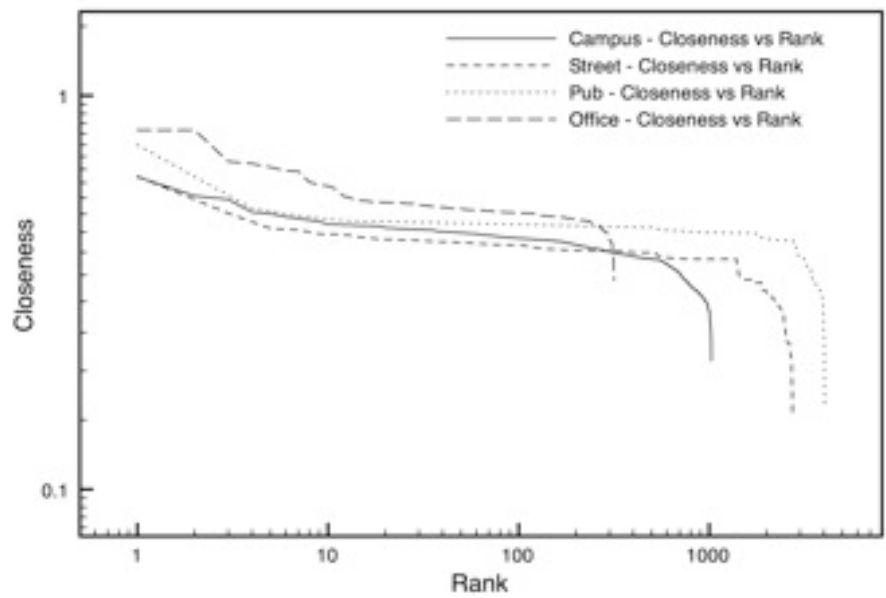

**Figure 5**: Ranked log-log plots of closeness for each of our four locations.

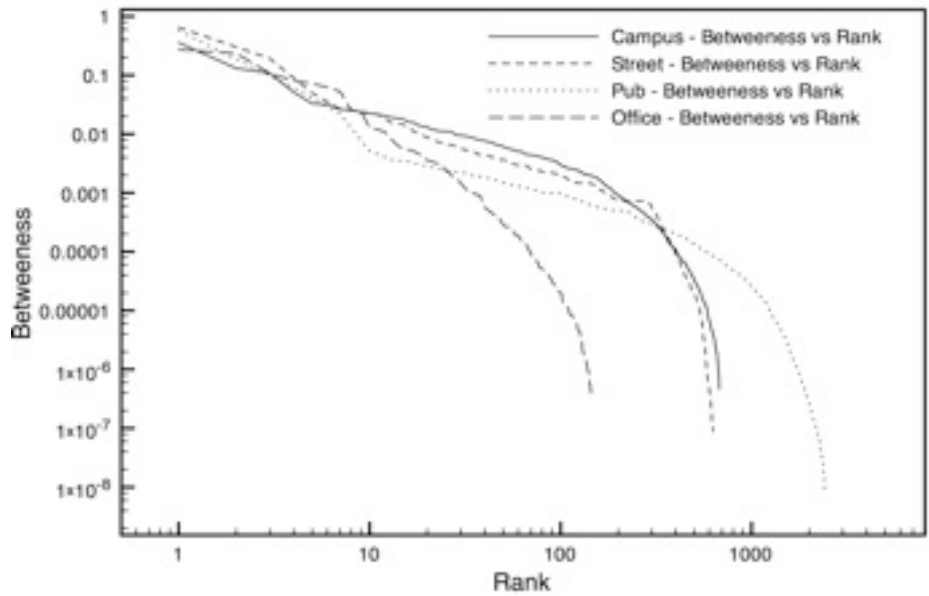

**Figure 6**: Ranked log-log plots of betweenness for each of our four locations.

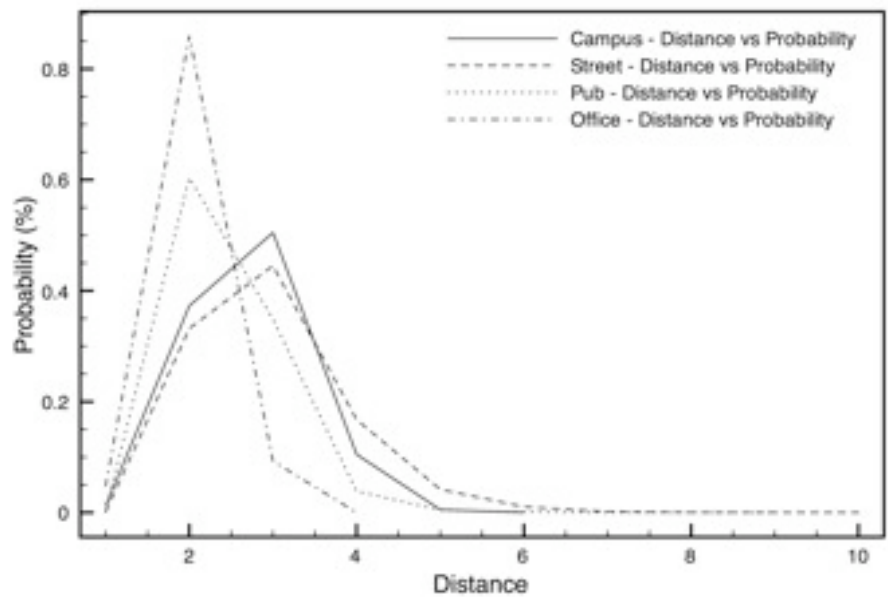

**Figure 7**: Probability plots of shortest path distance for each of our four locations.

When considering the whole range of values, degree graphs are overall more close to a power law distribution. Closeness graphs have short sharp tails, with a body that approximates a power law extremely well. Similarly, betweenness graphs have long sharp tails, while their body approximates a power law. The distance probability graphs can be approximated by a Poisson distribution.

The graphs we derived from analysing our Bluetooth data point to power-law distributions ($\gamma \approx 0.6$-$1.1$ for degree, $\gamma \approx 1.2$-$1.4$ for betweenness, $\gamma \approx 0.1$ for closeness) that are characteristic of scale-free, or self-similar networks. Such networks imply infinite variance, and usually in such networks there are a few nodes with extremely large number of links. Barabási et al. (1999a) have dubbed such networks 'scale-free', by analogy with fractals, phase transitions and other situations where power laws arise and no single characteristic scale can be defined. These characteristics can be found in kinship networks, physical and biological systems, and economic systems.

Scale-free networks have stimulated a great deal of theorising. The earliest work is due to [Simon, 1955], independently rediscovered by Barabási et al. [1999a; 1999b]. They show that scale-free networks emerge automatically from a stochastic growth model in which new nodes are added continuously and attach themselves preferentially to existing nodes, with probability proportional to the degree of the target node. Effectively, the richly connected nodes get richer.

We believe that our scanners recorded a phenomenon and process which is quite similar to the "rich getting richer" model, which explains the presence of power laws in our data. In terms of encounters, those people who have more links and encounters are the ones who are present more in the environment. When a new person comes along, chances are that they are going to encounter the regular customers or the employees. Thus, they share an encounter with an already well-connected person in the graph. It is this exact process that has been shown to result in power-law distributions.

## 6. Suitability for detention facilities

Our analysis suggests that intricate social processes were captured by our prototype, given adequate time and a large enough sample of people to be observed. Additionally, the underlying properties of our data suggest that our prototype did not capture noise, but somewhat of a "slice of reality". Interpreting the numbers generated by our algorithms can yield insight, but doing so requires knowledge of the scanning locations and the people being observed in them.

Similarly, analysing data captured by our system in a detention facility requires knowledge of the exact locations where the system was installed, as well as knowledge of the underlying behaviour of people in those areas. Analysts with such knowledge can draw on they automated data collection capability of our system to augment their ongoing efforts in understanding the links between various terrorists and organised crime organisations. There are, however, a number of issues that need to be resolved before utilising such a system for intelligence gathering in a detention facility.

To begin with, Bluetooth is only one possible technology that may be used for our purposes. Other proximity technologies such as RFID, NFC, and possibly ZigBee are all potential candidates for such a system. In fact, RFID would be the preferred mechanism, as RFID tags can easily be embedded in clothes or any other items that detainees may carry / be forced to carry. Key to the success of this scheme is the ability to relate each detainee to an individual or a set of RFID/Bluetooth identifiers. Ideally, these identifiers would persist for each individual across detention facilities.

An obvious issue with intentionally tagging individuals in a detention facility has to do with human rights abuse. While as technology developers we are merely highlighting the technological possibilities, we do wish to point out that our tagging system simply augments already established mechanisms of detention facilities, such as CCTV and human observation. Our system simply makes the identification of an individual detainee much quicker and more efficient, when compared to the analysis of days' worth of CCTV footage.

In addition to establishing the technological components required to deploy our system, an appropriate infrastructure is necessary so that data generated by our system can be readily accessed and analysed by intelligence agencies. This is most efficiently achieved by establishing a centralised data server, which will be used to store data arriving from various detention facilities. Subsequently intelligence agencies can issue queries to the data server, and retrieve the necessary information.

While in this paper we have presented a palette of tools and methods for analysing our systems data, further tools will be required in order to meet intelligence agencies' requirements. Ideally, our system will be used to augment ongoing investigations, by providing analysts with information that can be evaluated on a per-case basis. For instance, our system's central server could provide information about two people's relationship during their stay at a detention facility. Additionally, our system can provide an assessment of a suspect's social network, and people they are likely to contact once they are released from the detention facility. Effectively, analysts can look for patterns, or deviation from patterns in the data captured by our system, and evaluate these on an ad-hoc basis.

## 7. Conclusion and ongoing work

In this paper we describe our attempts to measure and quantify longitudinal socialising processes in a detention facility. We present a study where four distinct civilian locations were chosen for installing Bluetooth scanners which monitor the presence, and thus encounter, of people in those spaces. Our scanners generated a very rich data set that we used to derive social graphs for each of the four locations.

In our analysis we focused on the derived social graphs, and were able to compare various well-established properties and measurements of social graphs across the four locations. We found that the graphs exhibit power-law distributions when plotting their properties in rank-ordered graphs. These are characteristic of scale-free networks that can be found in kinship networks, physical and biological systems, and economic systems.

Our findings suggest that the utilisation of our system for capturing the socialising processes within detention facilities is a quire realistic strategy. This will require a number of issues to first be clarified, including the technological and infrastructure details, as well as the ethical and human rights challenges intrinsic to tracking and monitoring inmates.

As part of our ongoing work we are interested in exploring further our data sets. For example, we are interested in experimenting with different rules for generating the social graphs from the Bluetooth data. Also, we are in the process of running emulations of our data to explore ways in which information is diffused and spreads across the social networks.